# TALI: Protein Structure Alignment Using Backbone Torsion Angles

**Xijiang Miao**
Computer Science and Engineering
University of South Carolina
Columbia, SC, USA

**Michael G. Bryson**
Computer Science and Engineering
University of South Carolina
Columbia, SC, USA

**Homayoun Valafar**
Computer Science and Engineering
University of South Carolina
Columbia, SC, USA

***Abstract*— This article introduces a novel protein structure alignment method (named TALI) based on protein backbone torsion angle instead of the more traditional distance matrix. Because the structural alignment of the two proteins is based on comparison of two sequences of numbers (backbone torsion angles), we can take advantage of a large number of well developed methods such as Smith-Waterman or Needleman-Wunsch.

Here we report the result of TALI in comparison to other structure alignment methods such as DALI, CE and SSM ass well as sequence alignment based on PSI-BLAST. TALI demonstrated great success over all other methods in application to challenging proteins. TALI was more successful in recognizing remote structural homology. TALI also demonstrated an ability to identify structural homology between two proteins where the structural difference was due to a rotation of internal domains by nearly 180°.

**Index Terms— Protein structure; structural similarity; Molecular evolution; Torsion Angle; Ramachandran Space;**

## I. Introduction

The number of known protein structures is expanding in an exponential fashion. At the current rate, the PDB [1] database size will exceed 100,000 structures by the end of this decade. With the dazzling accumulation of protein structures, sophisticated and systematic analysis based on structural information becomes more and more feasible. Evolutionary relationship of organisms based on protein structures, protein function and protein-protein interaction can be cited as some examples of these types of analyses. Establishing the structural diversity among the cataloged library of structures can be viewed as the foundation of all of these analyses. Therefore the first step in analysis of protein structures is the establishment of structural relationship (similarity) between any two given proteins. In that regard, invaluable efforts have been made to classify and organize the protein structures, such as DALI (FSSP) [2, 3], CATH [4], and SCOP[5].

Several methods such as SSAP[6], DALI, CE[7], MAMMOTH[8] and SSM[9] have been developed to recognize the common core structural elements between protein candidates. Whether based on secondary structure configuration, distance matrix, or a pipeline of different procedures; these methods emphasize the overall structural similarity while neglecting the detail structural information.

Here we present a new method of structure alignment named TALI (Torsion Angle Alignment) that utilizes backbone torsion angles instead of a matrix of distances. The properties and statistics of torsion angles are now routinely used for protein quality check [10] by observing the distribution of the ($\varphi$, $\psi$) otherwise known as the Ramachandran plot [11]. The backbone structure of any protein can be represented by two major methods: a matrix of distances between designated atoms along the backbone (such $C_\alpha$) or the backbone torsion angles. Notably, the distance matrix is often times sparsely populated in the diagonal region and small islands corresponding to regions in close spatial vicinity. On the other hand, the torsion angle ($\varphi$, $\psi$) representation uses a series of angles that specifies the spatial relationship between peptide planes. Under theoretical conditions, a given list of torsion angles can be utilized to reliably reconstruct the backbone structure of a protein. However, a slight change in torsion angles can result a significantly different structure. The sensitive relationship between torsion angles and structure can enable us to perform detailed structural comparisons.

## II. Methods

### A. Structural distance based on torsion angles

The critical component of our implementation consists of implementation of an appropriate measure of structural relationship based on torsion angles. In the absence of statistical data defining the evolutionary importance of divergence in torsion angles, distance between any two given pairs of torsion angles need to be computed numerically. A large number of definitions can be adapted for computing the distance between two points in a *N*-dimensional space. Our first implementation relied on the most simplistic measurement of distance denoted as the "Plain TALI score" as shown in equations 1 and 2. Here $\varphi_{a,i}$ is the dihedral angle between the backbone C-N-Ca-C atoms of residue $i$ of protein $a$. $d_{ij}$ is the torsion angle distance measured by the Euclidean distance between two pairs of torsion angles ($\varphi$, $\psi$) (only matched pairs are counted). $N$ is the total length of aligned substructures.

Xijiang Miao is a Ph.D. student in the Department of Computer Science and Engineering at the University of South Carolina, Columbia, SC 29208.
Michael G. Bryson is a Ph.D. student in the Department of Computer Science and Engineering at the University of South Carolina, Columbia, SC 29208.
Dr. Homayoun Valafar is an assistant professor in the Department of Computer Science and Engineering at the University of South Carolina, Columbia, SC 29208.

The quantities $|\varphi_{a,i} - \varphi_{b,j}|$ and $|\psi_{a,i} - \psi_{b,j}|$ should be less than 180°, otherwise the 360° complementary angle should be used.

$$d_{ij} = \begin{cases} \sqrt{(\varphi_{a,i} - \varphi_{b,j})^2 + (\psi_{a,i} - \psi_{b,j})^2}, & \forall a_i \equiv b_j \\ NA & ,otherwise \end{cases} \quad ()$$

$$TS_p = \sqrt{\frac{1}{N}\sum_{ij} d_{ij}^2} \quad ()$$

Plain TALI score $TS_p$ treats all torsion angle changes equally. For example, a change within the α-helical region of Ramachandran space or a change from helical to beta strand region is not penalized differently. Albeit the results obtained based on this simplistic metric were very encouraging, here we do not report these results and focus on a more meaningful measure of distance.

The mentioned insensitivities of the trivial scoring mechanism can be remedied by redefining distances based on statistical information available from the Ramachandran plot. This addition will incorporate empirical likelihood of observing a particular pair of torsion angles. Here we define a function *R(l)*, which returns the minus log density at two points $(\varphi, \psi)_i$, $(\varphi, \psi)_j$ in Ramachandran space. The distance between any $(\varphi, \psi)_i$, $(\varphi, \psi)_j$ pair is defined in equation 3.

$$d_{ij}^r = \int_L R(l)dl \quad ()$$

Where $L$ is a straight path connecting two points $(\varphi, \psi)_i$ and $(\varphi, \psi)_j$. $d_{ij}^r$ represents the integration of torsion angle density along the path $L$. Intuitively, this Ramachandran distance analogues to measuring the total height of all steps in hill climbing with fixed step length. The final score summarizing the similarity between two pairs of torsion angles $TS_R$ is defined as shown in equations 4 and 5:

$$D_{ij} = \begin{cases} \exp(d_{ij}^r / 100) & \forall a_i \equiv b_j \\ NA & ,otherwise \end{cases} \quad ()$$

$$TS_R = \frac{1}{N}\sum_{ij} D_{ij} \quad ()$$

The *exp(d)* function shown in Eq (4), serves to heavily penalize larger distances which influence $TS_R$. The exponentiation term can be viewed as transition energy required migrating from one point in the Ramachandran space to another. This is a prelude to our future direction in implementation of distances based on Boltzmann distribution of energies between the two source and destination geometries. This algorithm can be greatly accelerated by pre-calculating the distances between discrete torsion angles.

### *B. Search for optimal structure alignment*

Our formulation of the structural alignment problem is conceptually very similar to that of the sequence alignment. While a traditional sequence alignment considers two sequences of characters, our approach considers alignment of two sequences of two dimensional numbers (torsion angles). Therefore, we have adapted a generalized Smith-Waterman algorithm [12] to find the best alignment using minimum $TS_R$ score in the place of PAM or BLOSUM substitution matrices.

### *C. Phylogeny tree generation*

Given a set of remote homologous protein domains, their evolutionary relationship can be inferred from $TS_R$. A distance matrix with $TS_R$ is generated for every pair of protein domain using the torsion angle alignment. Neighbor joining (NJ) [13] tree is applied on the distance matrix. The additive property of the distance is not guaranteed in structural alignment. Therefore the NJ tree only loosely reflects the homologous relationship between protein domains.

## III. Results

We have utilized results from SSM, DALI and CE as representatives of distance based methods aligning protein structures for comparison reasons. The results of DALI and CE are not shown since they are quite similar to SSM in the test cases reported here. In addition, PSI-BLAST [14] has been selected to provide sequence based alignment.

### *A. Alignment of protein structures using TALI*

This section illustrates the alignment result of protein 1nj8 [15] chain D and 1b76 [16] chain A by torsion angle alignment without referring to sequence or distance information. The percentage of sequence identity between 1nj8d and 1b76a is 17.5% using global alignment. The quality is checked against PSI-BLAST [14] and SSM [9]. PSI-BLAST collects position specific scoring matrix (PSSM) from a previous BLAST search result. The iterative process enables PSI-BLAST to find more distant relationships. SSM is based on matching graphs of protein's secondary structure elements, followed by an iterative 3D spatial alignment of protein backbone $C_\alpha$ atoms based on inter-atomic distances. The alignment result is good enough to verify the test cases used in this paper. Results reported by DALI and CE are also checked for quality evaluation.

The alignment 1nj8d and 1b76a is carried out independently by PSI-BLAST, TALI, and SSM as shown in Figure 1. The alignment result is coded by BioEdit. In this figure, the degree of amino acid similarity (or identity) is illustrated by a shaded box enclosing the aligned pair of amino acids (similarity measured based on PAM250). Larger boxes spanning regions of the two sequences indicate structural alignment. Solid box denotes confirmation of the alignment by at least another method while an open box denotes a one amino acid shift in the alignment frame. PSI-BLAST is performed through NCBI online query with default parameter settings. TALI uses gap open cost 600, gap extension 200, and Ramachandran distance as described in method section. SSM data is collected through EBI online server with default parameter settings, that is, match connectivity and normal precision.

Generally speaking, these three methods agree quite well with each other based on sequence, torsion angle, and distance strategy. PSI-BLAST is limited in identifying only 29% alignment based on sequence profile, which is in agreement with the results reported by SSM and TALI. TALI and SSM easily report larger aligned regions with similar degree of agreement. One curious outcome of TALI is that some alignment segments experience a “shift” in comparison to the results of SSM. This phenomenon happens in regions where the torsion angles are similar to each other (for example an α-helix) which is composed of torsion angles around (-65.3°,-39.4°) [17]. The quality of alignment can be improved by incorporating secondary structural information, physicochemical properties of each residue or even sequence similarity matrix such as PAM250. Please refer discussion section for explanation in detail.

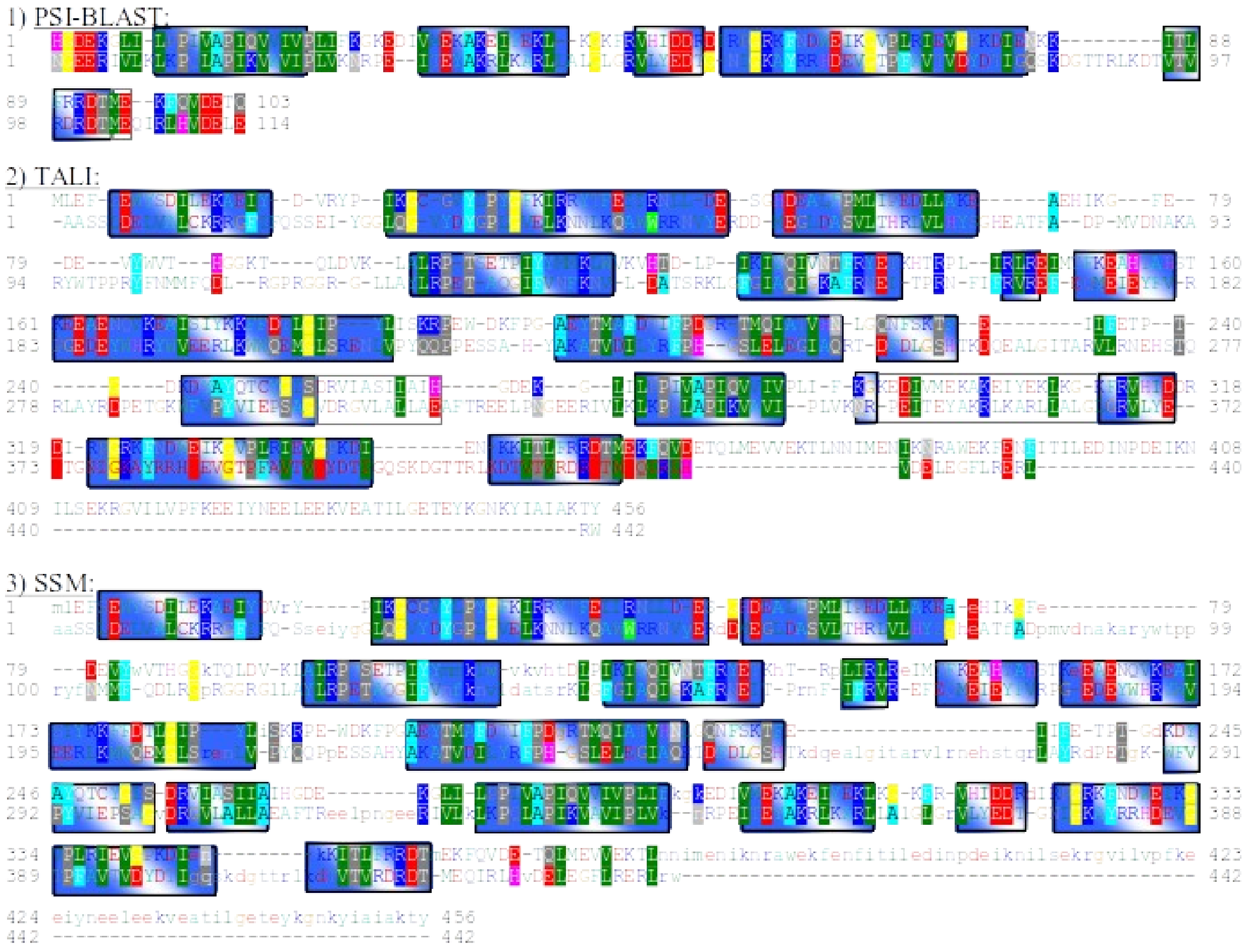

Figure 1 Align 1nj8D and 1b76A with PSI-BLAST, TALI, and SSM. Text with background indicates identical or similar amino acid. Solid box indicates confirmed alignment by at least two of the three algorithms; open box indicates one amino acid shift. 1) NCBI PSI-BLAST result with default parameters. 2) TALI result with gap open cost = 600, extension cost = 200, Ramachandran distance. 3) SSM result using default value. i.e., match connectivity and normal precision. Coding of amino acid similarity is generated by BioEdit.

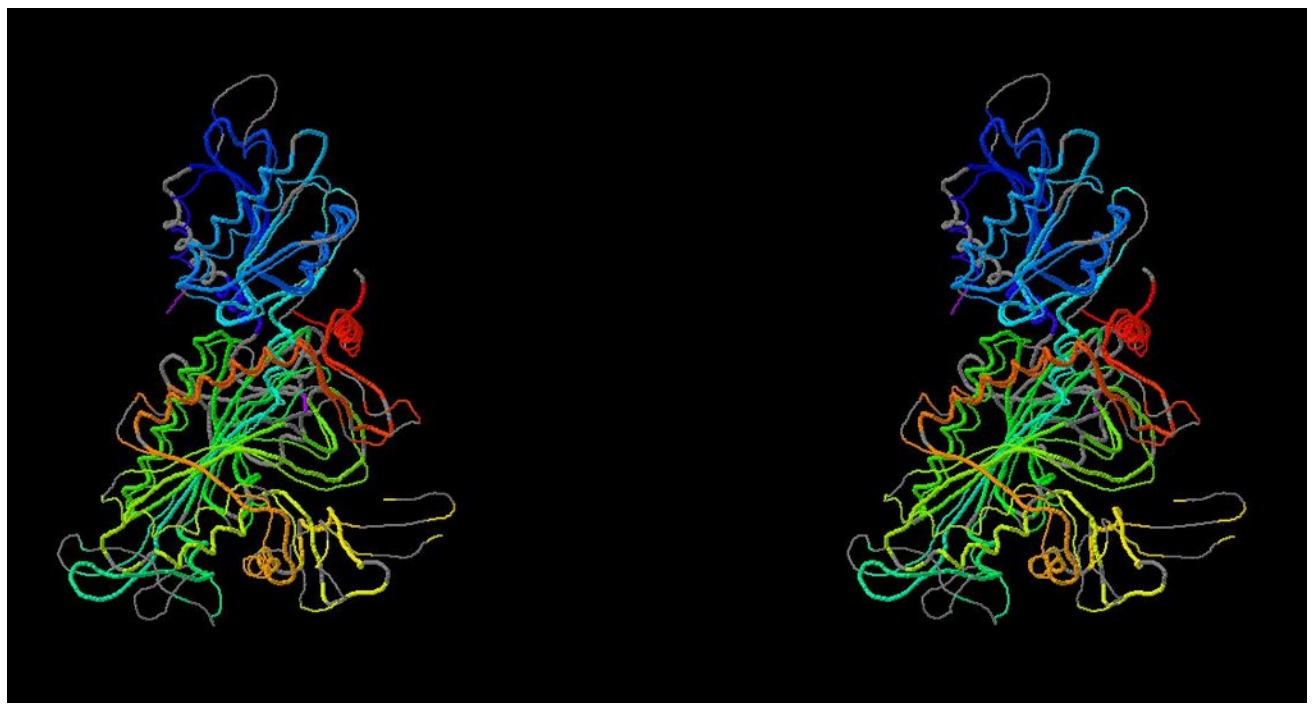

Figure 2 3D representation of 1nj8d and 1b76a alignment by TALI. The thicker trace is for 1nj8d. Threading is from red to purple. Gray is for gap. RMSD=7.79. RMSD at 75% quintile = 3.62. (Picture is prepared by Jmol)

The 3D representation of alignment is shown in Figure 2. The two proteins are correctly oriented according to the alignment result of DALI. Although the visual effect is satisfying, the root mean square deviation, or RMSD, is as high as 7.79 compared with 2.57 reported by SSM (matched pairs = 341). As shown in Figure 3, the distribution of distances between aligned residues is highly skewed. Simply dropping the aligned residues with farthest distances will greatly improve the RMSD with little sacrifice of number of matched residues. For example, by dropping residue pairs associated with top 25% largest distances, the RMSD can be dropped 3.62. Reorientation of the proteins based on the reduced alignment, RMSD can be further improved. Although RMSD is considered to be an important metric for alignment quality, it is proper only when the protein is assumed to be rigid.

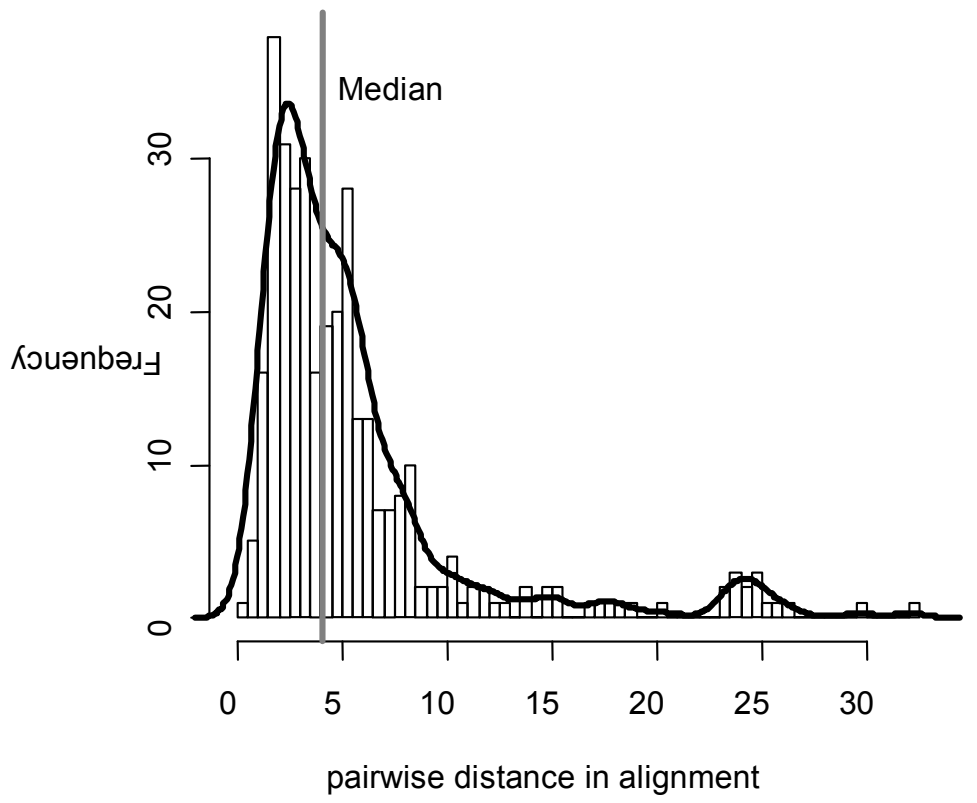

Figure 3 Histogram and distribution of pairwise distances in the alignment. Median of distances is 4.09. Initial RMSD=7.79. By dropping distances > 4.09, RMSD becomes 2.55; dropping top 25% distances (> 6.19), RMSD becomes 3.62. Total number of matched pairs is 334.

### B. *Performance of pure torsion angle alignment*

Fischer *et al.* [18] have presented 68 pairs of sequence-structure alignment test cases with different percentages of sequence identity and difficulty index for benchmarking sequence-structure alignment algorithm.

Figure 4 shows the alignment result of 1cewi [19] against 1mola [20], which is one of the top 10 difficult pairs among the list in [18]. 1cew and 1mol both are proteinase inhibitor with alpha-beta structure. For TALI, it is difficult to exactly align the structure because the torsion angles at the beta hairpin regions can not provide enough featured patterns to make a unique reliable alignment. This problem can be remedied by implementation of more analysis features discussed in the discussion section.

As a closely related example with obvious structural homology, 1cewi and 1r4ca [21] show that TALI outperforms SSM by correctly aligning the whole protein domain. Protein human cystatin C (HCC) inhibits papin-like cystein protease. Oligomerization of HCC leads to amyloid deposits. The N-terminal truncated variance of HCC (THCC, pdb code 1r4c) lacks the first 10 amino acid residues of the native sequence. The aggregation of THCC takes place through a domain swapping process. According to CATH [4] v2.6 classification, both 1cew and 1r4c belong to the sequence family 3.10.450.10.2 (proteinase inhibitor, cysteine). The protein sequence similarity is 44%, which strongly suggests their homologous origin. As shown in Figure 4, structure alignment from TALI correctly recovered the whole sequence alignment. The solid box shows the parts discovered by both TALI and SSM; dotted box shows the parts with high sequence similarity omitted by SSM but captured by TALI. From Figure 5 (a), we can conclude that one beta hairpin structure acts like a hinge. The remaining two subunits linked by this hinge remain relatively stable. DALI and CE show similar alignment results with SSM. Observation of mere distance relationship will miss the featured pattern in different subunit. The concept of protein domain conceptually divided the whole protein into small relatively independent units, which will help improve the quality of distance based alignment. But the protein domain boundary does not always have a unique clear cutoff [22]. Like sequence alignment, TALI does not rely much on the domain boundary definition; a properly defined protein chain is enough for alignment.

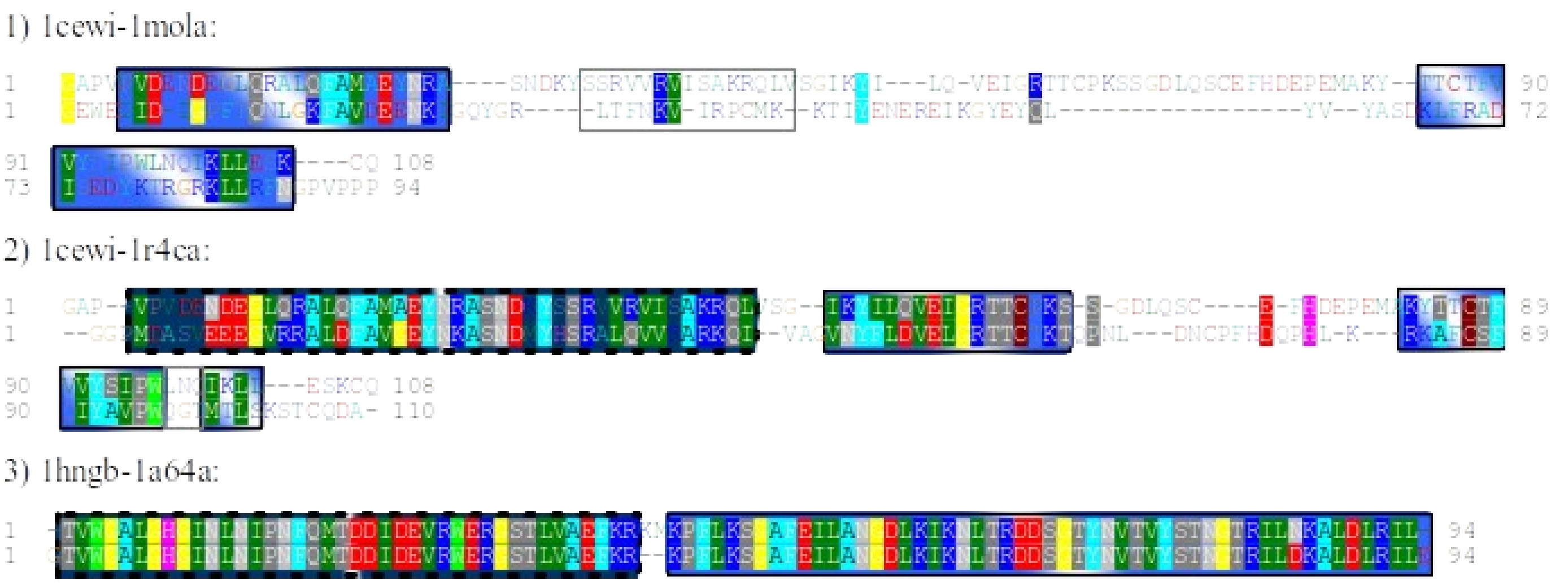

Figure 4 Three pairs of small proteins show pros and cons of torsion angle alignment. 1) 1cewi and 1mola. 2) 1cewi and 1r4ca. It is worth noting that they have more sequence identity. 3)1hngb and 1a64a. The protein sequences are almost identical. The 3D representation is shown in Figure 5.

Figure 5 (b) shows an extreme example, 1hngb [23] and 1a64a [24]. The amino-terminal domain of cell adhesion module CD2 (PDB id 1hng) can fold as monomer or metastable dimer. Murray et al. [24] engineered the protein by introducing a hinge-deletion mutant, which mimic the spontaneous random mutation during molecular evolution. The structural alignment by TALI correctly recovers the sequence alignment, while SSM does not find the first part of the structure (shown in dotted box in Figure 4 (3), 3D alignment is shown in Figure 5(b)). DALI or CE can not recover whole regions either.

Comparing the alignment result with DALI fold classification, we can often observe some examples that TALI performs better. For example, 1q59a [25] and 1g5ma [26] are related to BCL-2 family, while 1aa7a and 1mdta are put together by DALI but without any known biological support.

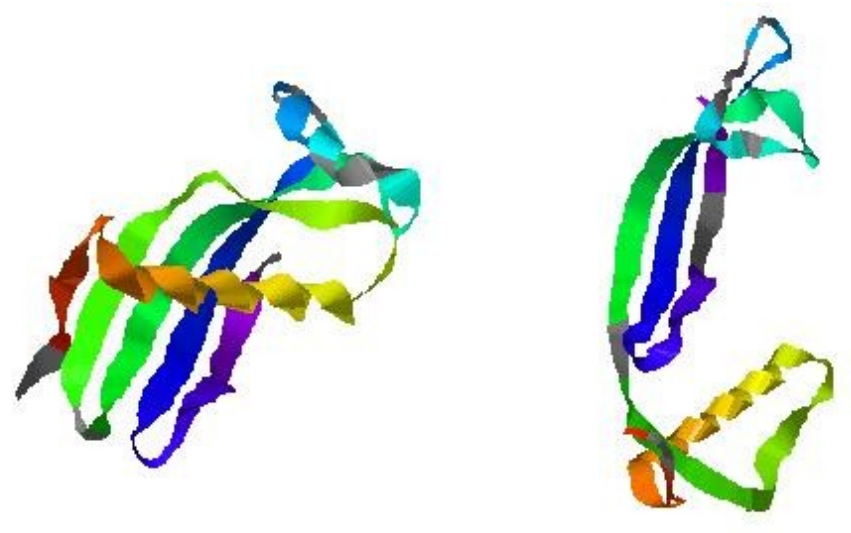

(a) 1cewi, 1r4ca

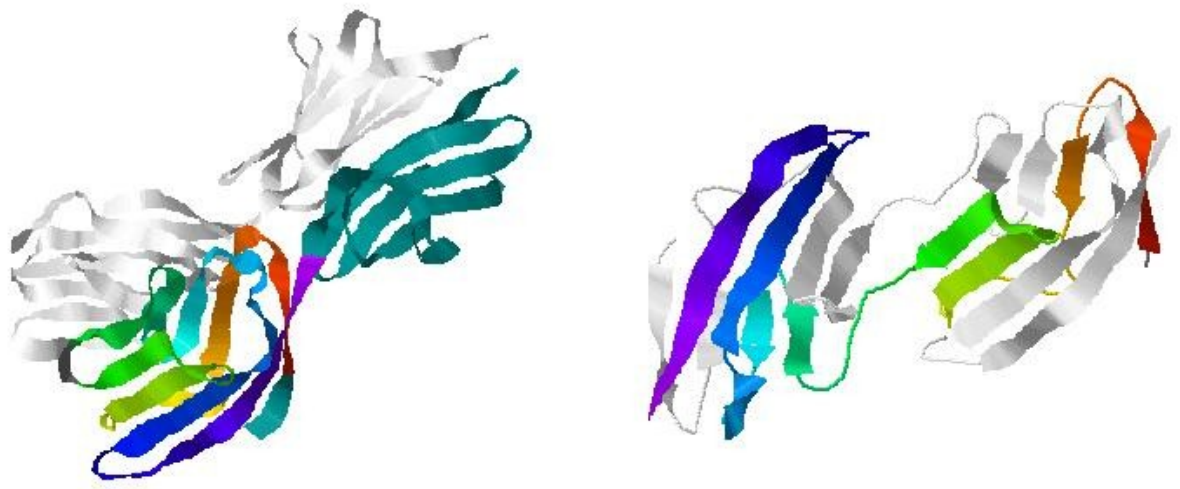

(b) 1hngb, 1a64a

Figure 5 3D representation of alignment of 1cewi-1r4ca and 1hngb-1a64a. In both cases, the beta hairpin structure acts like a hinge. The variance form is generated by open the hinge. The detail structure of the two parts remains relatively stable. Gray stands for gaps and cyan is for gaps at end, which are not counted in the final score.

### *C. Infer phylogeny from Torsion angle alignment*

To demonstrate the possibility of recovering phylogeny using TALI algorithm, we choose the protein phylogeny discussed by O'Donoghue and Schulten [27]. We use chain rather than domain as the taxa of the phylogeny. Use the chain information can avoid the artifact of domain boundary definition. And it will also be interesting to know the evolution history of specific chains, which is expected to be similar to the result of [27].

Figure 6 shows the result of TALI phylogeny for class II amonoacyl-tRNA synthetase, where the cost of opening a gap is set to 600 and extension cost is set to 200. The original tree appearing as FIG 10 and FIG 12 of the supplementary material in [27]. Each sub-class is shown by branch with thick lines.

Comparing the phylogeny tree derived by TALI and the original tree, the overall structure agree with each other quite well. For structures within subclasses, the difference is very subtle. For example, in branch $D_b$, the relationship is

(1eqrb, (1eqrc, (1eqra, (1c0aa, 1il2a))))

in TALI, while [27] gives

((1eqrb, 1eqrc), (1eqra, (1c0aa, 1il2a))).

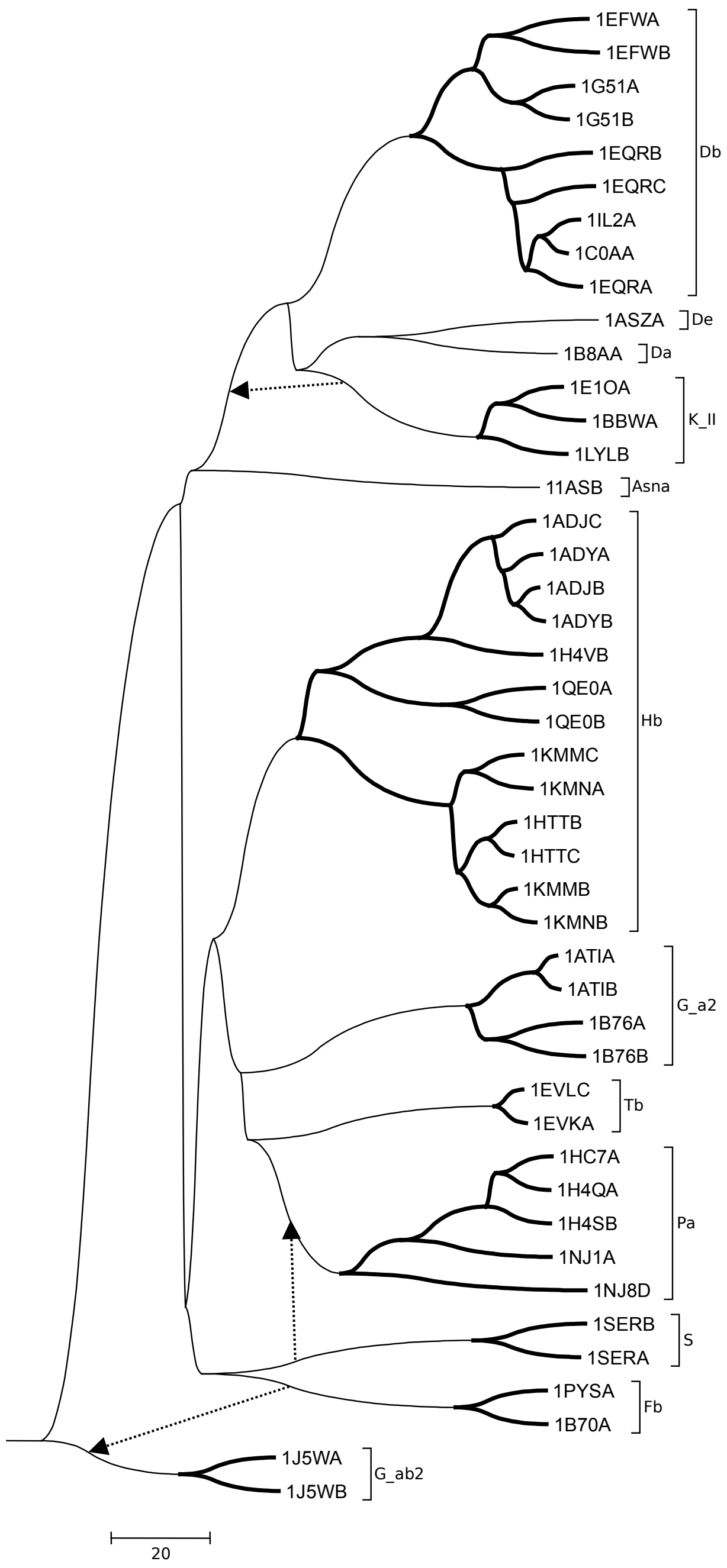

Figure 6 Full structural dendrogram of class II amonoacyl-tRNA synthetase. The chains are taken from supplement material FIG 2 of [27]. Neighbor joining (NJ) tree inferred from TALI score matrix. Tree is prepared by Mega3 [28].

Because the sequence identity is relatively high at subclass level, the disagreement can be resolved by using sequence alignment.

For the relationship between classes, the only differences lie in the position of $K_{II}$, S and $F_b$. The corresponding position is marked by dashed arrows. There is no miss-placed protein chain between sub-classes.

## IV. Discussion

### *A. TALI can align complex protein structures based on torsion angle information alone*

Torsion angles ($\varphi$, $\psi$) carry the majority of the backbone structural information. Although far more parameters are required to precisely define the relative position of atoms of protein backbone, most of them can be treated to be constant. For example, the dihedral angle of $\omega$ can be assumed to be 180° with standard deviation 5.8°, which maintains the planarity of the peptide plane. Other parameters include bond lengths, protein backbone angles, and so on.

While a list of torsion angles may not be sufficient in precise reconstruction of the protein backbone due to the accumulation of error, the torsion angle alignment is immune to this problem since alignment of the torsion angles occurs locally without a consideration of their relationship to distant amino acids (in the sequence). As a consequence, mutation of a few amino acids that would normally dramatically change the distance relationship could preserve most of torsion angle relationship. In this case, torsion angle can align the protein structures quite well, as shown in the listed in Figures 5 (a) and (b).

### *B. Improvement of TALI*

Through the course of structure alignment based on backbone torsion angles, some "shift" in alignment may occur due to low structural complexity of the region. We propose an adaptation of the following strategies to mitigate this effect:

1) Landmarks can be set at coil regions to guide the pure torsion angle alignment. This special treatment of torsion angles at coil region will make it focus on small segments which will increase the precision of alignment.
2) Some physicochemical properties of each residue can be used to enrich the information used for alignment. Incorporating sequence information is very natural for TALI. Except at the end of a polypeptide, there is exactly one pair of torsion angles ($\varphi$, $\psi$) per residue. The additional information per residue can be used as higher dimensional information added on the top of the torsion angles which will help capture more detail pattern and make a better alignment. Here our alignment of structure can use the results of sequence alignment to further refine structural alignment. Because the sequence alignment is well established for phylogeny analysis, the torsion angle alignment can readily borrow information from the result of sequence alignment, which is very useful to find the link among homologous proteins. Moreover, torsion angle alignment can also directly adapt methods from sequence alignment to extract out more meaningful structure information. For example, the torsion angle "motif" can be derived from a multiple alignment of torsion angles of homologous protein, which is expected to be a general extension of basic secondary structure elements (SSE) of protein, namely, α-helix and β-strands.

### *C. Evolutionary analysis using TALI*

The mutation and natural selection make proteins of offspring different from their ancestor in terms of sequence, structure and function. Homologous proteins and nucleotide sequences lose their similarity as time goes by in an exponential fashion. For example, according to Jukes-Canter model [29], nucleotide sequences of length $N$ are expected to have $\frac{3}{4}(1-e^{-4\alpha T})\times N$ sites with observable change after $T$ units of time past the divergence point (with substitution rate $\alpha$). If the sequence identity is only 25%~30%, or within the "twilight zone", it will be difficult to reliably estimate the protein evolutionary history using sequence comparison alone. The protein structure is considered to be more reliable in this case.

Protein structure similarity is considered to have two driving forces: physicochemical constrain and conservative evolution [30]. Zeldovich and his colleges [31] show that the physical constraint will drive different randomly generated sequences to same or very similar folds during simulated mutagenesis process; the sizes of these stable folds or "wonderfolds" are highly uneven, which is consistent with the uneven size of protein superfamily. The physicochemical induced protein structural similarity suggests that reliable protein homologue should be detected through sequence or redundant structure, which is not likely to be reproduced by chance in molecular evolution.

To infer phylogeny, the method of extract common core of structural similarity and make comparison is not accurate. Only compare the conserved cores are misleading due to two reasons:

1) The core is usually very stable and strongly selected, which leads to a converged evolution. That is, structures with different sources and function can have similar structure.
2) The structure comparison overlooks some detail structures to make a better match, which can be used as a source of homology detection. So this conserved core region is not an ideal material to study evolution.